\documentclass[aps,prl,floats,twocolumn,showpacs,superscriptaddress]{revtex4}

\usepackage{graphicx,epsfig}
\usepackage{times}
\usepackage{graphics,dcolumn,bm,fleqn,epic,eepic,float}
\usepackage{amssymb,amsmath,multirow,rotate,color}

\begin{document}

[Physical Review Letters {\bf 98}, 108103 (2007)]

\title{Dynamical Organization of Cooperation in Complex Topologies}

\author{J. G\'{o}mez-Garde\~{n}es}

\affiliation{Institute for Biocomputation and Physics of Complex
Systems (BIFI), University of Zaragoza, Zaragoza 50009, Spain}

\affiliation{Departamento de F\'{\i}sica de la Materia Condensada,
University of Zaragoza, Zaragoza E-50009, Spain}

\author{M. Campillo}

\affiliation{Departamento de F\'{\i}sica de la Materia Condensada,
University of Zaragoza, Zaragoza E-50009, Spain}

\author{L. M. Flor\'{\i}a}

\affiliation{Institute for Biocomputation and Physics of Complex
Systems (BIFI), University of Zaragoza, Zaragoza 50009, Spain}

\affiliation{Departamento de F\'{\i}sica de la Materia Condensada,
University of Zaragoza, Zaragoza E-50009, Spain}

\author{Y. Moreno}

\email{yamir@unizar.es}

\affiliation{Institute for Biocomputation and Physics of Complex
Systems (BIFI), University of Zaragoza, Zaragoza 50009, Spain}

\date{\today}

\begin{abstract}

In this Letter, we study how cooperation is organized in complex
topologies by analyzing the evolutionary (replicator) dynamics of the
Prisoner's Dilemma, a two-players game with two available strategies,
defection and cooperation, whose payoff matrix favors defection. We
show that, asymptotically, the population is partitioned into three
subsets: individuals that always cooperate ({\em pure cooperators}),
always defect ({\em pure defectors}) and those that intermittently
change their strategy. In fact the size of the latter set is the
biggest for a wide range of the "stimulus to defect" parameter. While
in homogeneous random graphs pure cooperators are grouped into several
clusters, in heterogeneous scale-free (SF) networks they always form a
single cluster containing the most connected individuals (hubs). Our
results give further insights into why cooperation in SF networks is
favored.

\end{abstract}


\maketitle

To understand the observed survival of cooperation among unrelated
individuals in social communities when selfish actions provide a
higher benefit, a lot of attention is being paid to the analysis of
evolutionary dynamics of simple two-players games like the Prisoner's
Dilemma (PD). In this game individuals adopt one of the two available
strategies, cooperation or defection; both receive $R$ under mutual
cooperation and $P$ under mutual defection, while a cooperator
receives $S$ when confronted to a defector, which in turn receives
$T$, where $T>R>P>S$. Under these conditions it is better to defect,
regardless of the opponent strategy, and assuming that strategies are
allowed to spread within the population according to their payoffs
(replicator dynamics \cite{hofbauer,gintis}), the proportion of
cooperators asymptotically vanishes in a well-mixed population ({\em
i.e.}  when each agent interacts with all other agents).

If the well-mixed population hypothesis is abandoned, so that
individuals only interact with their neighbors in a social network,
several studies
\cite{nowak,pacheco,doebeli,hauert,abramson,maxi,szabo} have reported
the asymptotic survival of cooperation on different types of
networks. Notably, cooperation even dominates over defection in
heterogeneous, SF networks where the distribution density of local
connectivities follows a power law. In view of the accumulated
evidence \cite{newmanrev,yamirrep} that many social (as well as
technological, biological and other) networks are highly
heterogeneous, these results are highly relevant for the understanding
of the evolution of cooperation.

In this Letter, we are interested in exploring the roots of the
diverse behavior observed on top of different complex topologies and
in providing an explanation in terms of microscopic arguments. We have
analyzed in detail the microscopic structural aspects underlying the
differences in the evolution of cooperation in a one-parameter family
of networks interpolating between Barab\'asi-Albert (BA) \cite{bara}
and Erd\"os-R\'enyi (ER) graphs. As usual in recent studies
\cite{nowak,pacheco}, we choose the Prisoner's Dilemma payoffs as
$R=1$, $P=S=0$, and $T=b>1$, and implement the finite population
analogue of replicator dynamics \cite{pacheco}. At each time step $t$,
which represents one generation of the discrete evolutionary time,
each node $i$ in the network plays with all its neighbors and
accumulates the obtained payoffs, $P_i$. Then, the individuals, $i$,
update synchronously their strategies by picking up at random a
neighbor, $j$, and comparing their respective payoffs $P_i$ and
$P_j$. If $P_i > P_j$, nothing happens and $i$ keeps the same strategy
for the next generation. On the contrary, if $P_j > P_i$, with
probability $\Pi_{i\rightarrow j}=(P_j-P_i)/\text{max}\{k_i,k_j\}b$,
$i$ adopts the strategy of $j$ for the next round robin with its
neighbors \cite{pacheco}.

We have performed simulations for a population of $N$ individuals that
interact following the couplings dictated by the underlying graph. To
explore the structure and dynamics of cooperative behavior in
different topologies, we have made use of the model developed in
\cite{jesus}, which allows to smoothly pass from a BA network to a
random graph of the sort of ER networks by tuning a single parameter
$\alpha\in (0,1)$. We will restrict hereafter to these two limiting
cases (ER, $\alpha=1$, and BA, $\alpha=0$). The results obtained for
other values of $\alpha$ will be discussed elsewhere \cite{noi2}. We
advance that they are consistent with the picture described in what
follows.

The dynamics is implemented once the network is grown. At the
beginning, each individual of the population has the same
probability of adopting either of the two available strategies:
cooperation ($s_i=1$) or defection ($s_i=0$). We let the system
evolve for $5000$ generations and check whether or not the system
has reached a stationary state as given by the fraction, $c(t)$, of
individuals that are cooperators. We impose that this magnitude is
in equilibrium when, taken over a time window of $10^3$ additional
generations, the slope of $c(t)$ is smaller than $10^{-2}$
\cite{note2}. After such a defined transient time $t_0$, we let the
system evolve again for $10^4$ additional time steps, and measure
the magnitudes whose behavior is described in the following. All
simulations presented hereafter have been carried out for networks
made up of $4000$ nodes with $\langle k \rangle=4$ and results are
averaged over at least $10^3$ different realizations of the networks
and initial conditions \cite{robustness}.

The above procedure allows to scrutinize in depth the microscopic
temporal evolution of cooperation as well as to characterize how its
local patterns are formed. Individuals' strategies asymptotically
(i.e. $t>t_0$) follow three different behaviors. Let $P(x,t)$ be the
probability that an individual adopts the strategy $x$ at any time
$t>t_0$. We say that an element $i$ of the population is {\em pure
cooperator} if $P(s_i=1,t)=1$, i.e., it plays as cooperator in all
generations after the transient time.  Conversely, {\em pure
defectors}, are those individuals for which $P(s_i=0,t)=1$. A third
class is constituted by {\em fluctuating} individuals, that is, those
elements that alternatively spend some time as cooperators and some
time as defectors.

\begin{figure}[!tb]
\begin{center}
\epsfig{file=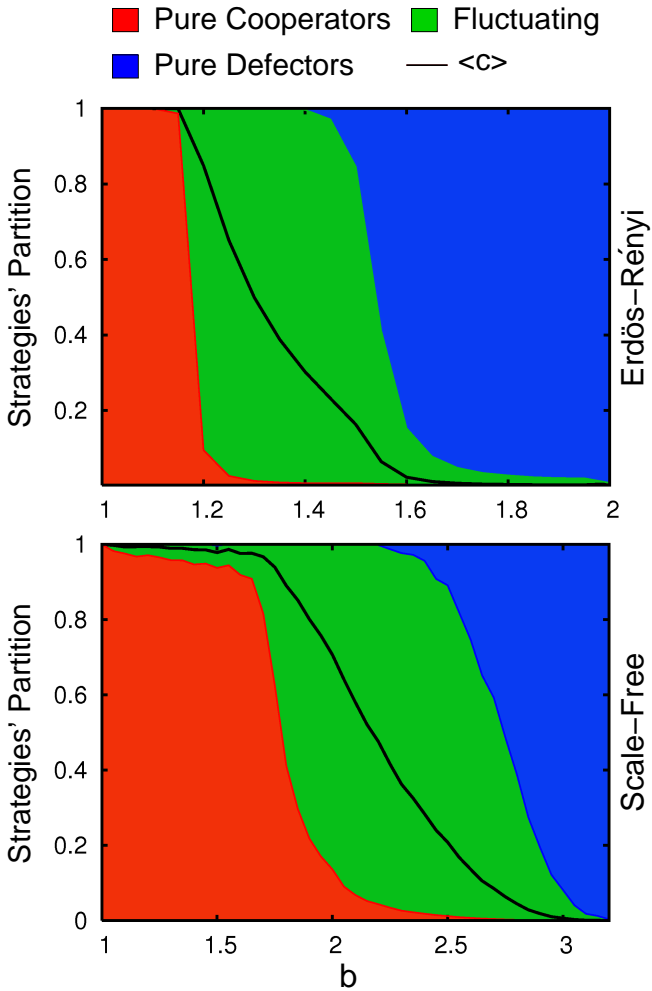,width=\columnwidth,angle=0}
\end{center}
\caption{(color online) Fraction (referred to the total number of
  individuals of the population) of pure and fluctuating strategies as
  a function of $b$. $\langle c\rangle$ (black continuous line)
  represents the asymptotic expected fraction of cooperators at each
  generation. The border lines separating colored regions correspond
  to $\rho_c(b)$ (red-green) and $1-\rho_d(b)$ (green-blue). Networks
  are made up of $4000$ nodes and $\langle k\rangle=4$. The exponent
  of the SF network is $-3$.}
\label{fig1}
\end{figure}

Figure\ \ref{fig1} shows the densities of the three classes of
players as $b$ is increased, for the two limiting cases of ER
(upper) and SF networks (bottom). Note that the fraction of pure
cooperators ($\rho_c$, continuous leftmost line) is always equal or
smaller than the density $\langle c\rangle(b)$, which is the
asymptotic expected value of the fraction of cooperators. This
indicates that the density of cooperators is, on average,
stationary, but not frozen as a significant fraction of individuals
are still able to intermittently adopt different strategies. It is
observed that in a small range of $b>1$, $\rho_c=\langle
c\rangle(b)$ for the ER network, while the equality does not hold
for any value of $b$ when the underlying architecture is a SF
network. Looking only at pure cooperation, there is a crossover for
moderate values of $b$. From that point on, the level of pure
cooperators in SF networks is above that in ER graphs. Additionally,
the decay of $\rho_c(b)$ is abrupt for homogeneous networks and more
smooth for SF ones. Therefore, pure cooperators are more robust to
variations of $b$ in these latter topologies.

\begin{figure}
\begin{center}
\epsfig{file=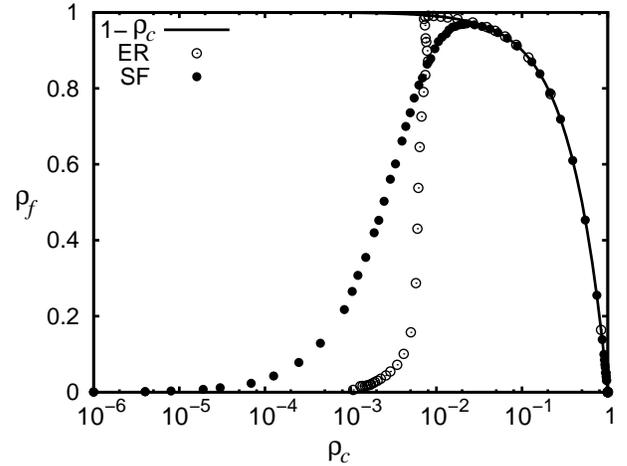,angle=0,width=\columnwidth}
\end{center}
  \caption{Fraction of fluctuating strategies as a function of the
  density of pure cooperators. The solid line is
  $\rho_f=1-\rho_c$. Deviations from it means that pure defectors have
  come into play. Note that the {\em same} value of $\rho_c$ in both
  topologies corresponds, in general, to different values of
  $b$. Networks parameters are those used in Fig.\ \ref{fig1}.}
\label{fig2}
\end{figure}

Furthermore, there is a region of $b$ in which almost {\em all}
strategies are fluctuating for the ER graph while this is not the case
for heterogeneous networks.  This feature is illustrated in Fig.\
\ref{fig2}, where it is represented the fraction of fluctuating
strategists, $\rho_f$, as a function of $\rho_c$. The deviation from
the continuous line, which is the function $\rho_f=1-\rho_c$, marks
the appearance of pure defectors. For both networks, the density of
fluctuating elements raises when $\rho_c$ decreases, however, the
decay of $\rho_f$ is clearly differentiated. While for the SF network
this magnitude falls smoothly and well below 1, for the ER network the
fraction $\rho_f$ continues to increase almost to $1$, and then
decreases suddenly, roughly keeping $\rho_c$ constant.  Moreover, the
number of pure cooperators relative to the total number of elements of
the population is significantly smaller in the ER networks than
in the SF case.

Figure\ \ref{fig2} gives even more information about what is going on
at a microscopic scale. Why does the fraction $\rho_c$ is smaller in
ER than in SF networks? An important clue comes from the analysis of the
local distribution of pure cooperators. Let us first define the
concept of cluster or core. A cooperator core ($CC$) is a connected
component (subgraph) {\em fully} and {\em permanently} occupied by
cooperator strategy $s_i=1$, i.e., by pure cooperators so that
$P(s_i(t)\neq 1, \forall t>t_0)=0$, $\forall i\in CC$. Analogously, a
defector core ($DC$) is the subgraph whose elements are pure
defectors, namely, when the condition $P(s_i(t)\neq 0, \forall
t>t_0)=0$, $\forall i\in DC$ is fulfilled. It is easy to see that a
$CC$ cannot be in direct contact with a $DC$, but with a cloud of
fluctuating elements that constitutes the frontier between these two
cores. Note that a $CC$ is {\em stable} if none of its elements has a
defector neighbor coupled to more than $k^c/b$ cooperators where $k^c$
is the number of cooperators linked to the element. Thus, the
stability of a $CC$ is clearly enhanced by a high number of
connections among pure cooperators, which implies abundance of cycles
in the $CC$.

\begin{figure}
\begin{center}
\epsfig{file=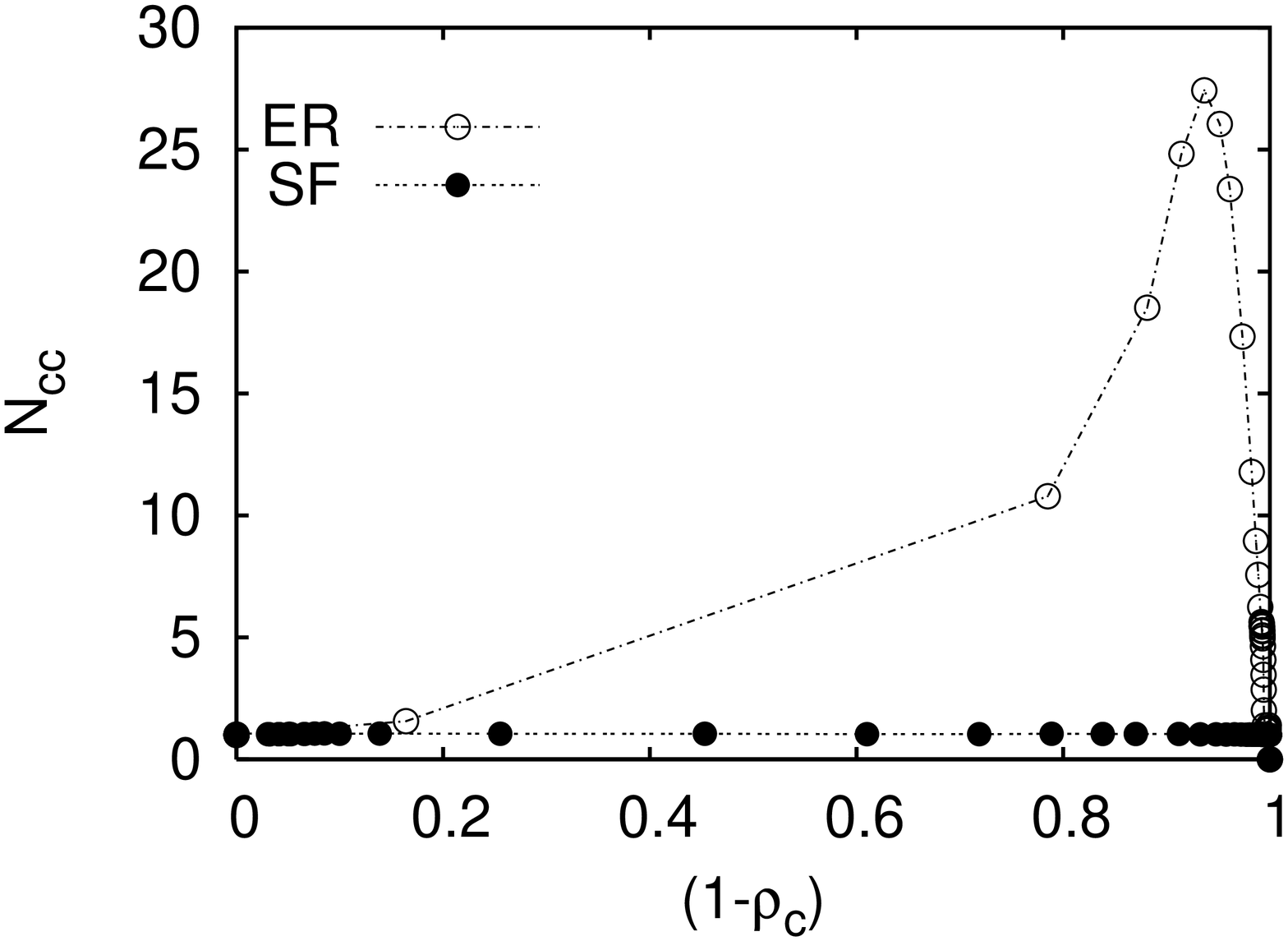,angle=-90,width=\columnwidth}
\epsfig{file=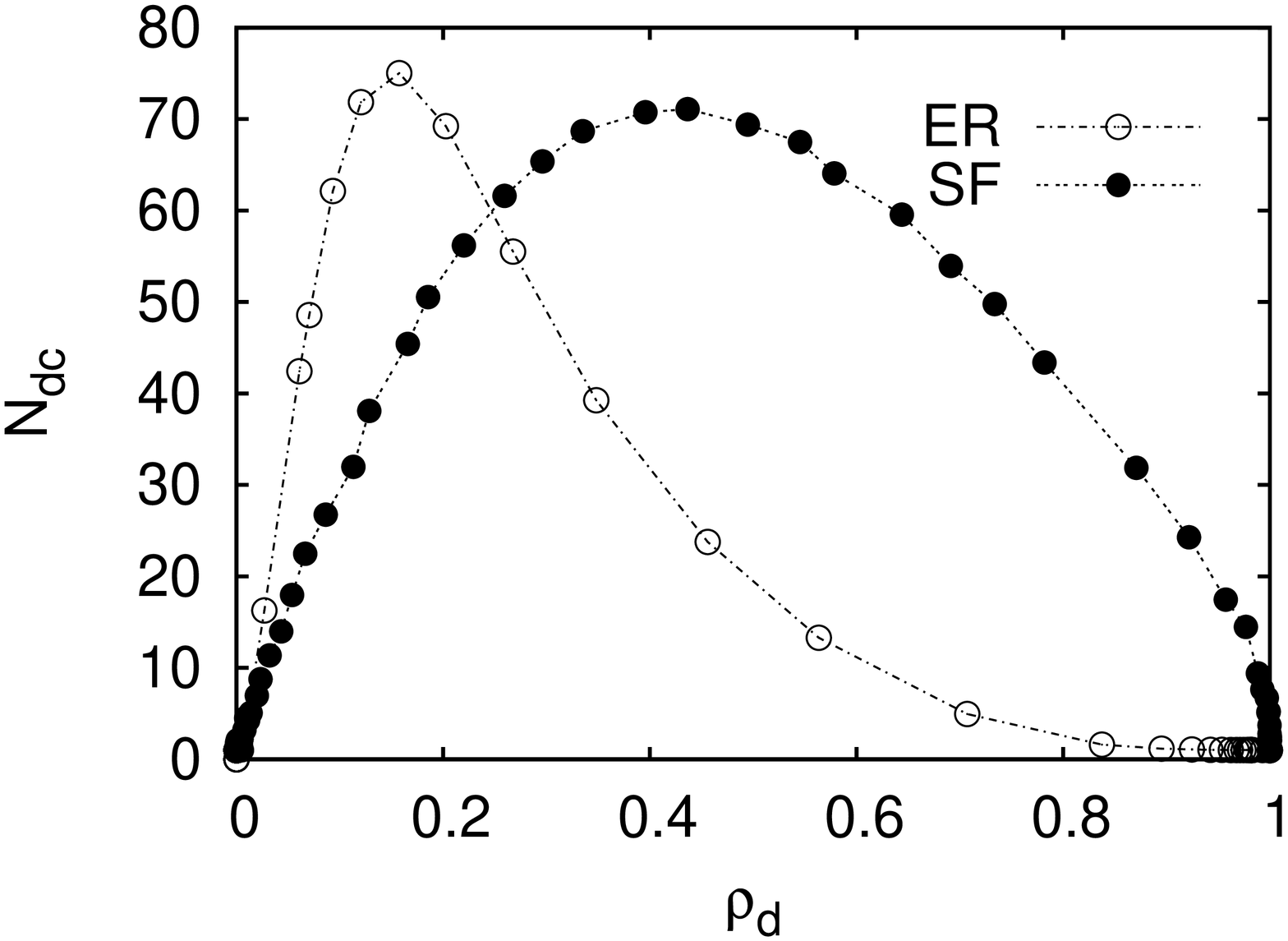,angle=-90,width=\columnwidth}
\end{center}
\caption{Number of clusters of pure cooperators (upper panel) and pure
  defectors (bottom panel) as a function of $1-\rho_c$ and $\rho_d$ in
  order to have both $x$ axes growing in the same way as $b$ does. The
  figures clearly show that it is possible to have more than one
  cooperator cluster only for the ER network, while pure defectors are
  always spread into several cores in SF networks and form a single
  cluster only in homogeneous structures. Network parameters are those
  of Fig.\ \ref{fig1}.}
\label{fig3}
\end{figure}

This microscopic structure of clusters is at the root of the
differences found in the levels of cooperation for both networks and
nicely explains why cooperative behavior is more successful in SF
networks than in homogeneous graphs. In fact, as far as
loops are concerned, the main difference between the two topologies is that
the number of small cycles of length $L$, $N_L$, are given by
\cite{gine2,enzo} $(\log(N))^L$ and $(\langle k\rangle -1)^L$,
respectively. Therefore, it is more likely that SF networks develop a
$CC$ than ER ones. This has been tested numerically by looking at the
probability that at least one cooperator core exists. The results
\cite{noi2} indicate that this probability remains $1$ for SF networks
even for $b>2$ and that it approaches zero for large $b$ following a
sort of second order phase transition. On the contrary, for ER
networks, the same probability departs from $1$ at an intermediate
value of $b$ and shows a sudden jump to $0$ at $b=2$, reminiscent of a
first order like phase transition.

We next focus on the detailed characterization of $CC$ and $DC$
structures. Fig.\ \ref{fig3} shows the number of clusters made up of
pure cooperators ($N_{cc}$, upper panel) and pure defectors ($N_{dc}$,
bottom panel) for both topologies as a function of $1-\rho_c$ and
$\rho_d$, respectively (recall that $1-\rho_c$ grows as $b$
increases). The first noticeable result concerns the number of
cooperator cores. While for ER networks $N_{cc}$ is equal to $1$ only
for a small range of $\rho_c$ values, and later increases up to a
maximum, for the SF network the number of such cores is always $1$, no
matter the value of $\rho_c$. That is to say, in one topology (ER),
there is a wide region of $b$ where there are several cooperator
cores, whereas pure cooperators in SF networks always form a single
core. On the other hand, the behavior of $N_{cc}$ in SF graphs implies
that the cycles discussed above are interconnected, giving rise to
more complex structures. We have also verified that the cooperator
core in SF networks contains the hubs, which are the ones that stick
together the cooperator cycles that would otherwise be disconnected
\cite{note}.

\begin{figure}
\begin{center}
\epsfig{file=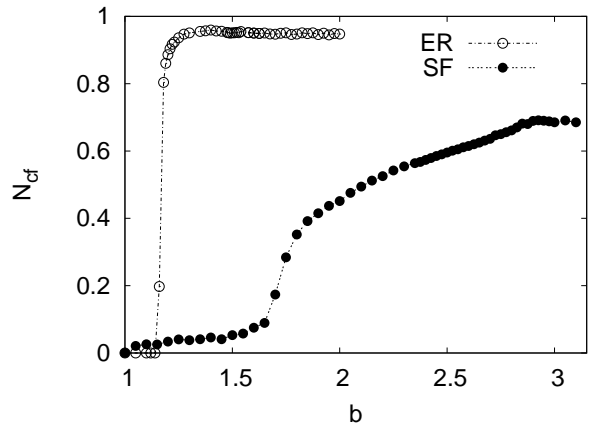,width=2.3in,angle=-90,clip=1}
\end{center}
  \caption{Fraction of pure cooperators that share at least a link
  with fluctuating strategists as a function of $b$. The high values
  of $N_{cf}$ for ER networks is influenced by the existence of
  several $CC$. This not the case for the SF network where there is
  only one $CC$ and $N_{cf}$ is much smaller. See the text for further
  details.}
\label{fig4}
\end{figure}

Looking again at Fig.\ \ref{fig3}, one realizes that there are also
radical differences in the local organization of pure
defectors. Again, the structural patterns in both networks can be
clearly distinguished. In ER networks, pure defectors first appear
distributed in several clusters that later coalesce to form a single
core for values of $b<2$, i.e., before the whole system is invaded by
defectors. Conversely, defectors are always organized in several
clusters for SF networks (except when they occupy the whole
system). This latter behavior results from the role hubs play
\cite{pacheco}. As they are the most robust against defector's
invasion, highly connected individuals survive as pure cooperators
until the fraction $\rho_c$ vanishes (see also Fig.\ \ref{fig2}), thus
keeping around them a highly robust cooperator core that loses more
and more elements of its outer layer until cooperation is finally
defeated by defection.

The picture emerging from the analyses performed clearly indicates
that two different paths characterize the raising (or breakdown) of
cooperation. This is also reflected in the way pure cooperators are
invaded by defector strategists. Starting at $b=1$ all individuals in
both topologies are playing as pure cooperators. However, for $b > 1$,
the pure cooperative level drops below $1$ and the population is
constituted by pure cooperators as well as by a cloud of fluctuating
individuals. When $b$ is further increased, these fluctuating
elements, that constitute the physical frontier of $CC$ clusters,
start to spend more and more time as defectors exploiting cooperators,
which ultimately leads to the invasion of the pure cooperators. In
this regard, the fragility of a $CC$ cluster mainly depends on how the
cluster's elements are exposed to their frontier. Figure\ \ref{fig4}
shows the "effective surface" of $CC$ as given by the fraction of pure
cooperators, $N_{cf}$, that share at least a link with fluctuating
players as a function of $b$. It is clear from the figure that in ER
networks the invasion is faster than in SF graphs, with a sudden
increase of $N_{cf}$ at low values of $b$, signaling the abrupt decay
observed in Fig.\ \ref{fig1}. This is due to the fragmentation of pure
cooperators into several $CC$ that are in a flood of fluctuating
elements, which eventually leaves pure cooperators more exposed to
invasion. Conversely, in SF networks, the existence of only one $CC$
makes it more resilient to defection since in this case the number of
pure cooperators exposed to fluctuating individuals is much lower, as
indicated by $N_{cf}$. Hence, the internal cohesion of the $CC$
cluster sustains cooperation for larger values of $b$ until an
$N$-defector core comes out. In summary, we have shown that there are
three different classes of individuals according to their asymptotic
strategies and that two different patterns of cooperative behavior,
determined by the underlying structure, can be clearly
identified. Specifically, our results unveil that in SF networks pure
cooperators always form a unique cooperator core while several
defector clusters coexist. On the contrary, for ER networks, both pure
and defector strategists are grouped into several clusters. The
microscopic organization of asymptotic cooperation on the evolutionary
dynamics of different games, like e.g. Snowdrift, Stag-Hunt
\cite{doebeli}, can obviously differ from the PD game. However, one
should expect that the partition of the population into fluctuating
and pure strategists will also generically hold. Finally, we note that
the same structural differences in the emergence and evolution of
cooperation has been pointed out in synchronization phenomena on top
of complex topologies \cite{noisync}. Whether or not these common
evolutionary patterns that emerge in two distinct phenomena are
relevant to explain the ubiquitous presence of SF networks in Nature
is still to be tested on more firm grounds. Studies of cooperation on
real social networks, like {\em e.g.} \cite{lozano} may help to scale
up to a mesoscopic description (in terms of communities) the
observations and results presented here.

\begin{acknowledgments}
  We thank A. Arenas, G. Bianconi, M. Marsili, and A. S\'{a}nchez for helpful
  comments and discussions. J.G.G. and Y.M. are supported by MEC
  through a FPU grant and the Ram\'{o}n y Cajal Program,
  respectively. This work has been partially supported by the Spanish
  DGICYT Projects FIS2004-05073-C04-01, and FIS2005-00337.
\end{acknowledgments}


\begin{thebibliography}{91}

\bibitem{hofbauer} J. Hofbauer and K. Sigmund, {\em Evolutionary Games and
Population dynamics}. (Cambrige University Press, Cambridge, UK, 1998).

\bibitem{gintis} H. Gintis, {\em Game Theory Evolving}. (Princeton
University Press, Princeton, NJ, 2000).

\bibitem{nowak} M.A. Nowak and R.M. May, Nature (London) {\bf 359}, 826
(1992).

\bibitem{pacheco} F.C. Santos and J. M. Pacheco, {\em Phys. Rev. Lett}
 {\bf 95}, 098104 (2005); F.C. Santos, J. F. Rodrigues and
 J. M. Pacheco, {\em Proc. Roy. Soc. Lond.} {\bf B 273}, 51 (2006).

\bibitem{doebeli} C. Hauert and M. Doebeli, Nature {\bf 428}, 643
  (2004); F. C. Santos J. M. Pacheco and T. Lenaerts,
  Proc. Nat. Acad. Sci. USA {\bf 103}, 3490 (2006).

\bibitem{hauert} H. Ohtsuki et al. 
, Nature {\bf 441}, 502 (2006).

\bibitem{abramson} G. Abramson, M. Kuperman, Phys. Rev. E., {\bf 63},
  030901(R) (2001); P. Holme et al, Phys. Rev. E., {\bf 68}, 030901(R)
  (2003).

\bibitem{maxi} V. M. Eguiluz et al.
, Am. J. Soc. {\bf 110}, 977 (2005).

\bibitem{szabo} G. Szab\'{o} and G. Fath, {\sc e-print cond-mat/0607344} (2006).

\bibitem{newmanrev} M. E. J. Newman, {\em SIAM Review \bf 45}, 167-256 (2003).

\bibitem{yamirrep} S. Boccaletti, V. Latora, Y. Moreno, M. Chavez, and D.-U.
  Hwang, {\em Phys. Rep. \bf 424}, 175-308 (2006).

\bibitem{bara} A. L. Barab\'{a}si, R. Albert, Science {\bf 286}, 509 (1999).

\bibitem{jesus} J. G\'{o}mez-Garde\~{n}es and Y. Moreno, {\em
Phys. Rev. E \bf 73}, 056124 (2006).

\bibitem{noi2} J. G\'{o}mez-Garde\~{n}es et al.,
in preparation.

\bibitem{note2} Note that a time window of $10^3$ time steps may not be enough
 to verify the condition for the slope of $c(t)$. In this case, we
 let the system evolve for as many time windows as needed, each one of $10^3$ generations.

\bibitem{robustness} We have checked that the results are robust for
  larger system sizes and do not depend on the time window
  used. Besides, they also hold for the PD parameterization,
  i.e., when $S=-\epsilon$, $\epsilon\ll 1$.


\bibitem{gine2} G. Bianconi and M. Marsili, J. Stat. Mech. P06005 (2005).

\bibitem{enzo} E. Marinari and R. Monasson, J. Stat. Mech. P09004 (2004).

\bibitem{note} The prominent role that the hubs play in the
maintenance of cooperation in SF networks has been correctly
emphasized in F. C. Santos and J. M. Pacheco, J. Evol. Biol. {\bf 19},
726 (2006).

\bibitem{noisync} J. G\'{o}mez-Garde\~{n}es, Y. Moreno, and A. Arenas,
  Phys. Rev. Lett. {\bf 98}, 034101 (2007).

\bibitem{lozano} S. Lozano, A. Arenas and A. Sanchez, in
preparation.

\end{thebibliography}
\end{document}